\documentclass[12pt,journal, onecolumn, draftcls]{IEEEtran}

\usepackage{bm,amsmath,array,booktabs}
\usepackage{cases}
\usepackage{wasysym}
\usepackage{amssymb}
\usepackage{amsfonts}
\usepackage{graphicx}
\usepackage{indentfirst}   
\usepackage{textcomp}
\usepackage{subfigure}
\usepackage{float}
\usepackage{multirow}
\usepackage{subeqnarray}
\usepackage{diagbox}
\usepackage{color}
\usepackage{url}

\newcommand{\pa}{\partial}

\begin{document}
	
\title{Transmit Array Interpolation for DOA Estimation via Tensor Decomposition in 2D MIMO Radar}

\author{Ming-Yang~Cao,~
	Sergiy~A. Vorobyov,~
	and Aboulnasr~Hassanien~
\thanks{This work was supported by the Academy of Finland (Grant No. 299243), National Nature Science Foundation of China (61171180);
	the fundamental Research Funds for Central Universities (Grant No. HIT. MKSTISP. 2016 13 and HIT. MKSTISP. 2016 26) and the Chinese scholarship council (CSC) under No. 201506120127. Parts of this paper were presented at IEEE ICASSP, Cancouver, BC, Canada, 2013 and IEEE SAM Workshop, Palma de Mallorca, Spain, 2016.}
\thanks{Ming-Yang Cao is with the Department of Communication and Signal Processing and Collaborative Innovation Center of Information Sensing and Understanding, Harbin Institute of Technology, China (e-mail:caomy@hit.edu.cn).}
\thanks{Sergiy A. Vorobyov is with the Department of Signal Processing and Acoustics,
	Aalto University, FI-00076 Aalto, Finland (e-mail: svor@ieee.org).}
\thanks{ Aboulnasr~Hassanien is with the Department of Electrical Engineering, Wright State University, Dayton, OH 45431, USA (e-mail:{hassanien}@ieee.org).}
\thanks{{\bf Corresponding author} is S.A. Vorobyov.}}
%

\maketitle

\begin{abstract}
In this paper, we propose a two-dimensional (2D) joint transmit array interpolation and beamspace design for planar array mono-static multiple-input-multiple-output (MIMO) radar for direction-of-arrival (DOA) estimation via tensor modeling. Our underlying idea is to map the transmit array to a desired array and suppress the transmit power outside the spatial sector of interest. In doing so, the signal-to-noise ratio is improved at the receive array. Then, we fold the received data along each dimension into a tensorial structure and apply tensor-based methods to obtain DOA estimates. In addition, we derive a close-form expression for DOA estimation bias caused by interpolation errors and argue for using a specially designed look-up table to compensate the bias. The corresponding Cram\'{e}r-Rao Bound (CRB) is also derived. Simulation results are provided to show the performance of the proposed method and compare its performance to CRB.
\end{abstract}


\section{Introduction}
The multiple-input-multiple-output (MIMO) radar, including the MIMO radar with transmit beamspace design, has received considerable attention during the last decade \cite{Fishler04c}-\cite{kha14}. Unlike phase-array radar, MIMO radar can simultaneously transmit several orthogonal waveforms through co-located antenna elements or widely separated antenna elements \cite{Li07}, \cite{Hai08}. Such waveform diversity brings many advantages to the MIMO radar over its phased-array radar counterpart among which are the possibility of flexible waveform design, improved parameter identifiability, higher spatial resolution, to name a few \cite{Fishler04c}-\cite{hass11}.  Moreover, MIMO radar has an inherent higher-dimensional structure that can be exploited to improve the estimation performance \cite{nion10}.

For a co-located MIMO radar, the transmit and receive antenna elements form particular array configurations, which provide a virtual array with increased number of antenna elements. This enables, for example, an improvement for direction-of-arrival (DOA) estimation performance. Many methods have been proposed to deal with the DOA estimation problem in MIMO radar, and most of them are transplanted from phased-array radar such as MUltiple SIgnal Classification (MUSIC), estimation of signal parameters via rotational invariance technique (ESPRIT) \cite{kha12}, maximum likelihood (ML) \cite{DC08}. Another class of DOA estimation methods exploits the higher-dimensional structure inherent in the MIMO radar \cite{nion10}. There are two main decomposition methods, parallel factor analysis (PARAFAC) \cite{Sidi00}, \cite{Kolda09} and higher-order singular value decomposition (HOSVD) \cite{Lath00}, \cite{Han14}. PARAFAC methods require a more relaxed uniqueness condition and have shown better estimation performance than covariance matrix singular value decomposition (SVD)-based methods. HOSVD utilizes the tensor structure, i.e., it iteratively performs SVD to every dimension of the received data. In the related R-dimensional harmonic retrieval (HR) parameter estimation problem, it is shown that the tensor-based methods have better DOA estimation performance and resolution ability than the matrix-based ones \cite{Haardt08}, \cite{Sun13}.

Both arrival and departure directions are often of interest, moreover, the transmit array or both the transmit and receive arrays are typically two-dimensional (2D). In the co-located MIMO radar, since the transmit waveforms are independently received by the receive array when transmitted from 2D array, the DOA estimation boils down to 2D (azimuth and elevation) parameter estimation. Several 2D estimation methods have been introduced in the context of MIMO radar \cite{jin09}, \cite{chan14}. Typically, a regular array geometries such as uniform rectangular array (URA) \cite{Zoubir12} or L-shaped array \cite{Liang10}, \cite{Xiaofei11} are assumed or other properties such as the existence of orthogonally polarized components of linear arrays \cite{Vershinin15} are used to reduce the search complexity while designing the DOA estimation techniques.  

The well known drawback of MIMO radar, which employs full waveform diversity, i.e., every transmit antenna element radiates orthogonal waveform, in the regime of parameter estimation is the reduction of the signal-to-noise ratio (SNR) gain comparing with phased-array radar at a fixed total transmit energy \cite{hass10}, \cite{hass11}. In order to improve the SNR gain in the known spatial sector of interest, coherent transmit beamspace (TB) design techniques have been developed \cite{hass11}, \cite{hass13}. These designs are called transmit beamspace MIMO radar, and they utilize the fact that the target spatial region is usually a priori known before solving the parameter estimation problem, thus the transmit beam could focus the transmit energy within the desired spatial region while suppressing the sidelobes outside of it. The TB MIMO radar takes the advantages of the phased-array and MIMO radars and enjoys both the waveform diversity and high SNR gain within the sector of interest. As a result, comparing to MIMO radar with full diversity, TB MIMO radar has better DOA estimation performance and lower Cram\'{e}r-Rao bound (CRB), moreover, the number of waveforms for achieving the best perfomance can be found optimally \cite{hass11}.

Another merit of TB MIMO radar is the possibility to reduce the DOA estimation computational burden \cite{hass132}-\cite{Cao16}. Specifically, for an arbitrary planar array MIMO radar, there are more than one parameter to be estimated such as elevation/azimuth, polarization, Doppler, etc.  Thus, grid searching methods, e.g., spectral MUSIC, become very time consuming or infeasible. A nonuniform transmit array structure, which does not lead to Vandermonde structure for the steering vectors along any of the dimensions, prohibits the application of computationally efficient methods based on polynomial rooting instead of a grid searching. The  TB design methods, however, allow to interpolate the transmit array into a virtual transmit array which has a desirable regular structure \cite{hass132}, \cite{Cao16}. The interpolation can be designed so that the signal subspace of the received snapshots will enjoy some desirable properties such as the rotational invariance property. In turn, it enables us to use polynomial rooting based parameter estimation techniques, which greatly reduces the computational burden even if the transmit array has an arbitrary planar structure. Note that different from the more traditional interpolation strategies applied used for receive array \cite{Fried92}, \cite{pes02}, the interpolation in the context of TB MIMO radar needs to be performed as the transmit array. In addition to the above mentioned drawbacks, MIMO radar usually suffers from pulses storing condition. Comparing with the number of pulses, MIMO radar has a relative large dimension, which makes the estimation of the sample covariance matrix infeasible. Furthermore, performing a singular value decomposition (SVD) to a large scale covariance matrix demands a high computational burden.  

In this paper, we develop a 2D transmit array interpolation for mono-static MIMO radar with irregular planar transmit array. We also develop a search-free 2D DOA estimation methods based on tensor modeling. To achieve high SNR gain within a known spatial sector/region of interest, we design a transmit array interpolation matrix limiting the interpolation errors between the actual steering vector and the desired one to a certain level within the sector of interest, while minimizing the sidelobes outside the sector. We have first reported the transmit array interpolation approach in \cite{hass13} for the case of mapping an irregular transmit array into an L-shaped array. Here, we generalize the virtual array structure to several practically appealing structures, which satisfy the translational invariance property. The transmit array interpolation optimization problems are also generalized to arbitrary norm formulation to allow for smaller interpolation errors and sidelobes. With proper transmit array interpolation matrix design, computationally efficient methods can be applied to the 2D TB MIMO radar DOA estimation with an arbitrary array configuration. It is worth noting that, in addition, we do not require the receive array calibration information. Then we propose a general tensor modeling for the 2D TB MIMO radar with transmit array interpolation that fully capitalizes on its multidimensional structure. Under the small number of pulses condition, we consider the deterministic MIMO radar model, i.e., direct data approach, which requires significantly less computational burden. By performing HOSVD to the received MIMO radar snapshots, we achieve the tensor-based signal subspace and then develop tensor-based methods to obtain DOA estimates. Since the interpolation errors are not negligible in our model, we analyze the DOA estimation bias caused by the interpolation errors under high SNR condition. In addition, we explain how to build an offline look-up table to decrease the bias. Furthermore, we derive the corresponding transmit array interpolation 2D TB MIMO radar deterministic CRB as a benchmark for verifying the performance of the proposed methods.

The paper is organized as follows. The basic notations and MIMO radar signal model are given in Section \uppercase\expandafter{\romannumeral2}. In Section \uppercase\expandafter{\romannumeral3}, we develop the 2D TB MIMO radar interpolation method and present the corresponding direct data TB MIMO radar model, while the tensor modeling of 2D TB MIMO radar as well as matrix and tensor based ESPRIT type elevation and azimuth estimation methods are developed in Section \uppercase\expandafter{\romannumeral4}. Section \uppercase\expandafter{\romannumeral5} is devoted to CRB derivation, the analysis of interpolation errors, and the development of a lookup table-based bias compensation technique in DOA estimation. Section \uppercase\expandafter{\romannumeral7} reports simulation results showing the performance of the proposed approach under a variety of conditions, and Section \uppercase\expandafter{\romannumeral8} summarizes our conclusions. Some technical derivations of a supporting nature are given in Appendices.

\section{Notation and Signal Model}
\subsection{Notation}
Scalar is denoted by italic letters $a$, column vector by lower-case bold letter $\bm{a}$, matrix by upper-case bold letter $\bm{A}$, tensor by calligraphic bold-face letter $\bm{\mathcal{A}}$, respectively. The $i$-element of a vector is denoted as $a_i$, the $(i_1, i_2)$-element of a matrix $\bm{A}$ is denoted as $a_{i_1, i_2}$, and the $(i_1, i_2, \ldots, i_N)$-element of an $N$-order tensor $\bm{\mathcal{A}} \in \mathbb{C}^{I_1 \times I_2 \times \ldots \times I_N}$ as $a_{i_1, i_2, \ldots, i_N}$.  We use the notations $(\cdot)^*, (\cdot)^T, (\cdot)^H, (\cdot)^{-1}, \circ, \odot, \otimes, \ast, \dagger$, and $\ddagger$ for denoting complex conjugate, transpose, Hermitian transpose,  matrix inverse, outer product, Khatri-Rao product, Kronecker product, Hadamard product, pseudo-inverse, and projection, respectively. In addition, $\| \cdot \|_p$ and $\| \cdot \|_{\rm F}$ stand for the $p$-norm of a vector and the Frobenius norm of a matrix, respectively, $\mathbb{E} \{ \cdot \}$ denoted the mathematical expectation, $\bm{I}_M$ is an $M\times M$ identity matrix, and $\text{diag}(\bm{a})$ denotes a diagonal matrix that holds the entries of $\bm{a}$ on its diagonal.

\textit{Definition 1} (The scalar product): The scalar product of two tensors $\bm{\mathcal{A}}$, $\bm{\mathcal{D}} \in \mathbb{C}^{I_1\times I_2\times\ldots I_N}$ is given by
\begin{equation}
c \triangleq <\bm{\mathcal{A}}, \bm{\mathcal{D}}> \stackrel{\triangle}{=} \sum_{i_1=1}^{I_1} \sum_{i_2=1}^{I_2} \dots\sum_{i_N=1}^{I_N} d_{i_1,i_2,\ldots,i_N}^*\cdot a_{i_1,i_2,\ldots,i_N} .
\end{equation}
Using this definition, the higher-order norm of a tensor is given by 
\begin{equation}
\|\bm{\mathcal{A}}\|_\text{H} \triangleq \sqrt{<\bm{\mathcal{A}},\bm{\mathcal{A}}>}.
\end{equation}

A matrix unfolding of a tensor $\bm{\mathcal{A}}$ along $I_n$th mode is denoted as $\bm{\mathcal{A}}_{(I_n)}$ following the notation in, for example, \cite{Haardt08}. We define the concatenation of two tensors along $I_n$th mode as $\bm{\mathcal{A}} \sqcup_{I_n} \bm{\mathcal{D}}.$ By fixing $I_n$th mode to a specific value $k$, we obtain a sub-tensor denoted by  $\bm{\mathcal{A}}_{I_n=k}$. Respectively, the vectorization of an $N$-order tensor $\bm{\mathcal{A}}$ along $I_n$th mode is defined as $\text{vec}(\bm{\mathcal{A}}) \triangleq \text{vec}(\bm{\mathcal{A}}_{(I_n)}^T)$. 

\textit{Definition 2} (The $n$-mode tensor-matrix product): The $n$-mode product of a tensor $\bm{\mathcal{A}}\in\mathbb{C}^{I_1\times I_2\times\cdots\times I_N}$ and a matrix $\bm{D}\in\mathbb{C}^{J \times I_n}$ along $n$th mode is given by
\begin{align}
&\bm{\mathcal{C}} \triangleq \bm{\mathcal{A}}\times_n\bm{D}  \\ \notag 
&c_{i_1, i_2, \ldots, i_{n-1}, j_n ,i_{n+1},\ldots,i_N} = \sum_{i_n=1}^{I_n} a_{i_1, i_2, \ldots, i_N} \cdot d_{j, i_n} \label{eq:nmode}
\end{align}
where $\bm{\mathcal{C}}\in\mathbb{C}^{I_1\times I_2\times\ldots\times I_{n-1}\times J\times I_{n+1}\times \ldots\times I_N}$.

\textit{Definition 3} (The outer product): The outer product of $N$-order tensor $\bm{\mathcal{A}}\in\mathbb{C}^{I_1\times I_2\times\ldots I_N}$ and $M$-order tensor $\bm{\mathcal{D}}\in\mathbb{C}^{J_1\times J_2\times\ldots J_M}$ is given by
\begin{align}
&\bm{\mathcal{C}} \triangleq \bm{\mathcal{A}}\circ\bm{D}\in\mathbb{C}^{I_1\times I_2\times\cdots \times I_N\times J_1\times J_2\times \cdots \times J_M}  \\ \notag 
&c_{i_1,i_2,\ldots,i_N,j_1,j_2,\ldots,j_M}=a_{i_1,i_2,\ldots,i_N}\cdot d_{j_1,j_2,\ldots,j_M} .
\end{align}

\subsection{MIMO Radar Signal Model}
Consider a mono-static MIMO radar system equipped with $M$ co-located transmit antenna elements and $N$ co-located receive antenna elements. The transmit and receive arrays are assumed to be placed on a plane and have arbitrary geometries. The receive array consists of antenna elements randomly selected from the transmit array.  The transmit antenna elements are assumed to be located at the position $\bm{p}_m \triangleq [x_m, \, y_m]^T,m=1,2,\ldots,M$.  Then the $M\times 1$ steering vector of the transmit array can be expressed as
\begin{equation}
	\bm{a}(\theta,\phi) \triangleq [e^{-j2\pi\bm{u}^T (\theta,\phi) \bm{p}_1}, \ldots, e^{-j2\pi\bm{u}^T (\theta,\phi) \bm{p}_M}]^T
\end{equation}
where $\bm{u}(\theta,\phi) \triangleq [\sin\theta\cos\phi ,\, \sin\theta\sin\phi]^T$ represents the propagation vector, and $\theta$, $\phi$ are the elevation and azimuth, respectively. Receive antenna elements are randomly chosen out of $M$ elements in the transmit array. Similarly, the steering vector of the receive array can be then expressed as 
\begin{equation}
	\bm{b}(\theta,\phi) \triangleq [e^{-j2\pi\bm{u}^T (\theta,\phi) \bm{p}_1}, \ldots, e^{-j2\pi\bm{u}^T(\theta,\phi)\bm{p}_N}]^T.
\end{equation}

Let $s_m(t)$ be the complex envelope of the $m$th transmit signal where $t$ represents the fast time, and then $\{ s_{m} (t) \}_{m = 1}^{M}$ be a set of $M$ waveforms. Each waveform  $s_m(t)$ has unit energy, and all waveforms are assumed to be orthogonal to each other during one pulse,  i.e., $\int_T s_m(t)s_{m^\prime}^*(t)dt=\delta(m-m^\prime)$, where $T$ is the radar pulse duration, $\delta(\cdot)$ denotes the Dirac delta function, and $L$ is the number of samples per pulse period. The signal radiated towards a spatial region of interest is therefore given by
\begin{equation}
	\zeta(t,\theta,\phi) =\bm{a}^T (\theta,\phi) \bm{s} (t) = \sum_{m=1}^{M} {a}_m(\theta,\phi) s_m(t) \label{eq:radiated signal} 
\end{equation}
where $\bm{s}(t) \triangleq [s_1(t),\ldots,s_M(t)]^T$ and ${a}_m(\theta,\phi)$ is the $m$th element of $\bm{a}(\theta,\phi)$. 

We assume that radar cross section (RCS) coefficients obey Swerling \uppercase\expandafter{\romannumeral2} model \cite{skol08}. It means that the RCS coefficients remain unchanged during one pulse while vary for different pulses. There are $K$ targets to be located in the spatial sector of interest.  Therefore, the received MIMO observation vector can be expressed as
\begin{equation}
	\bm{x}(t,q)=\sum_{k=1}^K\beta_k(q)(\bm{a}^T(\theta_k,\phi_k)\bm{s}(t))\bm{b}(\theta_k,\phi_k)+\bm{n}(t,q)
\end{equation}
where $q$ represent the slow time index, $\beta_k(q)$ is the RCS coefficient of $k$th target  with variance $\sigma_{\beta}^2$, and $\bm{n}(t,q)$ is the noise vector modeled as complex spatial and temporal white Gaussian process. Using the orthogonality property of the transmit waveforms,  the received data vector corresponding to the $m$th waveform after matched-filtering can be obtained as
\begin{equation}
	\bm{y}_m(q) = \sum_{k=1}^K \big( a_m (\theta_k,\phi_k) \bm{b} (\theta_k, \phi_k) \big) \beta_k(q) + \bm{z}(q)
\end{equation}
where $\bm{y}_m (q) \in \mathbb{C}^{N\times 1}$, $\bm{z}(q)$ is the noise vector after matched-filtering whose covariance matrix is given by $\sigma^2_n\bm{I}_M$. Hence, the whole receive vector, i.e., the vector that is obtained by stacking  $\bm{y}_m(q), \, m=1,\ldots,M$,  one under another, can be written as
\begin{equation}
	\bm{y}(q)=(\bm{A}(\theta,\phi)\odot\bm{B}(\theta,\phi))\bm{\beta}(q)+\bm{z}(q)   \label{eq:receive model}
\end{equation}
where $\bm{A}(\theta,\phi) \triangleq [ \bm{a} (\theta_1,\phi_1), \ldots, \bm{a} (\theta_K,\phi_K) ]$ is the transmit steering matrix, $\bm{B} (\theta,\phi) \triangleq [\bm{b} (\theta_1,\phi_1), \ldots, \bm{b} (\theta_K,\phi_K)]$ is the receive steering matrix, and $\bm{\beta}(q) \triangleq [\beta_1(q), \ldots, \beta_K(q)]^T$ is the vector of RCS coefficients during $q$th pulse.

\section{2D Transmit Array Interpolation}
In many real world applications, we only need to consider a particular spatial sector instead of the whole space. The incident target spatial sector is thus assumed hereafter to be a prior knowledge. Then we can propose a 2D transmit array interpolation design aiming to focus the transmit energy within a pre-designed spatial sector while suppressing the sidelobes outside the sector of interest. In addition, it is needed to interpolate the actual steering vector of an arbitrary array into a virtual one which has a desired structure. In doing so, during the DOA estimation, we should be able to benefit from the use of the virtual array so that computationally efficient techniques can be applied and/or the higher-dimensional structure inherent in the 2D TB MIMO radar model can be exploited. Note that in this way, the initial waveform design and transmit array interpolation problems are decoupled, and the initial waveforms are multiplied by the desired interpolation matrix before being radiated from the antenna elements of the actual transmit array.

\subsection{Desirable 2D Virtual Transmit Array Structures}
Let $\bm{E}\stackrel{\triangle}{=}[\bm{e}_1,\ldots,\bm{e}_{\tilde{M}}]$ denote a transmit beamspace and interpolation matrix, such that
\begin{equation} 
\bm{E}^H\bm{a}(\theta,\phi) = \bm{\tilde{a}}(\theta,\phi)  \label{eq:interpolation}
\end{equation}
i.e., $\bm{a}(\theta,\phi)$ is mapped to $\bm{\tilde{a}}(\theta,\phi)$ that has a desired structure. A design technique for such interpolation matrix $\bm{E}$ can be developed in the same manner as the 2D transmit beamspace techniques \cite{hass132}, while enforcing additionally the condition \eqref{eq:interpolation}. Here we discuss several desired structures of $\bm{\tilde{a}}(\theta,\phi)$.  

If we map the transmit array steering vector to a coordinate product, i.e., Khatri-Rao product, of two arrays with $M_1$ and $M_2$ elements, respectively, the steering vector $\bm{\tilde{a}}(\theta,\phi)$, designed with enforcing \eqref{eq:interpolation}, can be expressed as
\begin{equation}
\bm{\tilde{a}}(\theta,\phi)=\bm{c}(\mu) \odot \bm{d}(\nu) \label{eq:virtual}
\end{equation}
where 
\begin{align}
\bm{c} (\mu) \triangleq [1, e^{-j2\pi x_1\mu}, \ldots, e^{-j2\pi x_{(M_1-1)}\mu}]^T  \label{eq:carray} \\
\bm{d} (\nu) \triangleq [1, e^{-j2\pi y_1\nu}, \ldots, e^{-j2\pi y_{(M_2-1)}\nu}]^T. \label{eq:darray}
\end{align}
Here,  $\bm{c}(\mu)$ and $\bm{d}(\nu)$ represent the steering vectors of two virtual arrays, respectively. The parameters $\mu$ and $\nu$ in the steering vectors of the virtual arrays \eqref{eq:virtual}--\eqref{eq:darray} are generalized direction cosine functions. In fact, the virtual arrays can have different structures depending on how we set the generalized parameters $\mu$ and $\nu$. 

If we set $\mu=\sin\theta$ and $\nu=\sin\phi$, the virtual array \eqref{eq:virtual} can be described as a Khatri-Rao product of two perpendicular uniform linear arrays (ULAs) placed along X-axis and Y-axis. In this case, the steering vectors of the vertical and horizontal ULAs carry the elevation and azimuth, respectively. The desired inter-element spacing has to be at least half-wavelength to avoid the ambiguity in angle estimation. 

If we set $\mu=\sin\theta\cos\phi$ and $\nu=\sin\theta\sin\phi$, then the corresponding virtual array represents the URA.

If we map the transmit array into an L-shaped array, we have $\bm{\tilde{a}}(\theta,\phi) = [\bm{c}^T(\mu), \ \bm{d}^T(\nu)]^T$. Recalling \eqref{eq:carray} and \eqref{eq:darray}, the virtual array in this case can be described as an L-shaped array placed on X-axis and Y-axis, which contains $M_1+M_2-1$ elements, i.e., one element is common for both the vertical and horizontal arrays.

Finally, the number of virtual antenna elements $\tilde{M} \leq M$ depends on which virtual array we design. For virtual two perpendicular/cross ULAs, we have $\tilde{M}=M_1+M_2$; for virtual URA, we have $\tilde{M}=M_1M_2$; and $\tilde{M}=M_1+M_2-1$ for virtual L-shaped array.

\subsection{Transmit Array Interpolation By Minimax Optimization}
The spatial region of interest is the available prior knowledge Let $\theta_{g}\in\Theta, g=1, \ldots, G_{\theta}$ be the chosen angular grid which properly approximates the desired elevation sector $\Theta$ by a finite number $G_{\theta}$ of calibration points. Similarly, let $\phi_{g} \in \Phi, g = 1, \ldots, G_{\phi}$ be the  chosen angular grid  which properly approximates the desired azimuth sector $\Phi$ by a finite number $G_{\phi}$ calibration points. The interpolation matrix ${\bm{E}}$ can be calculated as the least squares (LS) solution to \eqref{eq:interpolation}. For this, we define the $M\times G_{\theta} \cdot G_{\phi}$ and the $\tilde{M}\times G_{\theta} \cdot G_{\phi}$ matrices ${\bm A}$ and $\tilde{\bm A}$,
respectively, as follows
\begin{equation}
\bm{A} (\Theta,\Phi) \triangleq [\bm{a}(\theta_1,\phi_1),\ldots, \bm{a}(\theta_{G_{\Theta}}, \phi_1),\ldots, \bm{a}(\theta_{G_{\Theta}}, \phi_{G_{\phi}})]
\end{equation}
\begin{equation}
\tilde{\bm{A}} (\Theta,\Phi) \triangleq [\tilde{\bm{a}}(\theta_1,\phi_1), \ldots,
\tilde{\bm{a}} (\theta_{G_{\Theta}}, \phi_1), \ldots, \tilde{\bm{a}} (\theta_{G_{\Theta}}, \phi_{G_{\phi}})].
\end{equation}
There are more calibration points than the antenna elements in the actual arrays ($G_{\theta} \cdot G_{\phi}\geq M$), thus $\bm{A}(\Theta,\Phi)$ and $\tilde{\bm{A}}(\Theta,\Phi)$ have full rank. The LS solution to \eqref{eq:interpolation} is given as
\begin{equation}\label{eq:LSsolution}
\bm{E}=\left( \bm{A}(\Theta,\Phi)\bm{A}^H(\Theta,\Phi) \right)^{-1} \bm{A}(\Theta,\Phi) \tilde{\bm{A}}^H(\Theta,\Phi).
\end{equation}
The LS solution \eqref{eq:LSsolution} provides the optimal approximation performance if we only concerned with minimizing the interpolation errors between the actual array and a desired one. However, for matrix $\bm{E}$ design in TB MIMO radar, LS solution \eqref{eq:LSsolution} does not enable controlling the sidelobe levels of the transmit beampattern. Therefore, the resulting sidelobe levels can be even higher than the in-sector levels. This may result in wasting most of the transmit power in the out-of-sector areas which can lead to severe performance degradation. 

To enable controlling sidelobe levels, we propose to use the minimax criterion to minimize the difference between the steering vectors of the actual array and the desired virtual array within the sector of interest, while keeping the sidelobe levels outside the spatial sector of interest bounded by a given maximum level. Therefore, the interpolation matrix design problem can be formulated as the minimax optimization problem as follows
\begin{eqnarray}
\!\!\!\!\!\!\!\!\!&\!\!\!\!\!\!&\!\!\! \min_{\bm{E}}\max_{\theta_{g},
\phi_{g^{\prime}}}  \left\| \bm{E}^H{\bm a} (\theta_{g}, \phi_{g^{\prime}})
- \tilde{\bm{a}}(\theta_{g}, \phi_{g^{\prime}}) \right\|_{p} \label{eq:OptObj_1}
\\
\!\!\!\!\!\!\!\!\!&\!\!\!\!\!\!&\!\!\! \theta_{g}\in\Theta,\ g = 1, \ldots, G_{\theta},\quad   \phi_{{g}^{\prime}}\in\Phi,\ g^{\prime} = 1,\ldots, G_{\phi} \nonumber \\
\!\!\!\!\!\!\!\!\!&\!\!\!\!\!\!&\!\!\!{\rm subject\ to}  \left\| {\bm{E}}^H{\bm a} (\theta_{h}, \phi_{h^{\prime}})  \right\|_{p} \leq \gamma, \label{eq:OptConst_1} \\
\!\!\!\!\!\!\!\!\!&\!\!\!\!\!\!&\!\!\! \theta_{h}\in\bar\Theta,\ h=1, \ldots, H_{\theta}, \quad  \phi_{h^{\prime}} \in \bar\Phi,\ h^{\prime} = 1, \ldots, H_{\phi} \nonumber
\end{eqnarray}
where $\bar\Theta$ and $\bar\Phi$ are the out-of-sector regions in the elevation and azimuth domains, respectively, $\theta_{h} \in \bar\Theta,\ h = 1, \ldots, H_{\theta}$ and
$\phi_{h^{\prime}} \in \bar\Phi,\ h^{\prime} = 1, \ldots, H_{\phi}$ are angular grids which properly approximate the out-of-sector regions $\bar\Theta$ and $\bar\Phi$, respectively, $\gamma$ is a positive number of user choice used to upper-bound the worst-case sidelobe level, and any combination of $l_1$-, $l_2$-, or $l_\infty$-norms can be used in the objective and constraints of \eqref{eq:OptObj_1}--\eqref{eq:OptConst_1}. It is worth noting that for $l_2$-norm, the optimization problem \eqref{eq:OptObj_1}--\eqref{eq:OptConst_1} becames a convex quadratically constrained quadratic programming (QCQP) problem. As such, it can be solved by interior-point methods \cite{Nesterov} implemented, for example, in SeDuMi solver \cite{cvx}, which is used in the well known CVX MATLAB package. However, QCQP problems in general have heavy computational burden. Moreover and more significant, flat sidelobes over the elevation and azimuth cannot be achieved by \eqref{eq:OptObj_1}--\eqref{eq:OptConst_1} in the case of $l_2$-norm. Instead, if $l_1$- and $l_\infty$-norms are used in \eqref{eq:OptObj_1}--\eqref{eq:OptConst_1}, the interpolation error for every value on the grid can be bounded, and sidelobes can be better controlled. In addition, for $l_1$- and $l_\infty$-norms, the optimization problem \eqref{eq:OptObj_1}--\eqref{eq:OptConst_1} can be cast as a linear programming (LP) problem, which can be efficiently solved by many existing LP solvers, including SeDuMi solver.

Alternatively, it is possible to minimize the worst-case out-of-sector sidelobe level while upper-bounding the norm of
the difference between the interpolated array steering vector and the desired one. This can be formulated as the following optimization problem
\begin{eqnarray}
\!\!\!\!\!\!\!\!\!&\!\!\!\!\!\!&\!\!\! \min_{\bm{E}} \max_{\theta_{h}, \phi_{h^{\prime}}} \left\| {\bm{E}}^H{\bm a}(\theta_{h},\phi_{h^{\prime}}) \right\|_{p} \label{eq:OptObj_2} \\
\!\!\!\!\!\!\!\!\!&\!\!\!\!\!\!&\!\!\! \theta_{h}\in\bar\Theta,\ n=1, \ldots, H_{\theta}, \quad  \phi_{h^{\prime}} \in \bar\Phi,\ h^{\prime} = 1, \ldots, H_{\phi} \nonumber \\
\!\!\!\!\!\!\!\!\!&\!\!\!\!\!\!&\!\!\!{\rm subject\ to} \left\| {\bm{E}}^H {\bm a} (\theta_{g}, \phi_{g^{\prime}}) - \tilde{\bm a}( \theta_{g}, \phi_{g^{\prime}}) \right\|_{p} \leq \Delta \label{eq:OptConst_2} \\
\!\!\!\!\!\!\!\!\!&\!\!\!\!\!\!&\!\!\! \theta_{g}\in\Theta,\ g =1, \ldots, G_{\theta}, \quad \phi_{g^{\prime}} \in \Phi,\ g^{\prime} = 1, \ldots, G_{\phi} \nonumber
\end{eqnarray}
where $\Delta$ is a positive number of user choice used to control the deviation of the interpolated array from the desired one, which is also called interpolation errors tolerance. 
However, the interpolation matrix design based on solving the above optimization problem will introduce interpolation errors to the virtual transmit array, which will cause a bias in the DOA estimation. The relationship between the DOA estimation performance and the interpolation errors will be discussed in Section~\uppercase\expandafter{\romannumeral6}. Finally,  note that the transmit array interpolation design does not involve waveform design. In fact, the problems of initial orthogonal waveforms design and transmit array interpolation are decoupled. Then the transmission process can be presented so that the initial orthogonal waveforms are first multiplied by the transmit beamspace and interpolation matrix $\bm E$ and then the resulting waveforms are radiated from the antenna element of the actual transmit array. In turn, the reception process remains standard, that is, matched filtering to the initial orthogonal waveforms.

\subsection{2D TB MIMO Radar Signal Model}
Similar to \eqref{eq:radiated signal}, the signal radiated in the 2D TB MIMO radar after transmit array interpolation \eqref{eq:interpolation} is given by 
\begin{align}
\tilde{\zeta} (t, \theta, \phi) &= \bm{\tilde{a}}^T (\theta,\phi) \tilde{\bm{s}} (t) = \left( \bm{E}^H \bm{a} (\theta,\phi) \right)^T \tilde{\bm{s}} (t) \notag \\
&= \bm{a}^T (\theta,\phi) \bm{E}^*  \tilde{\bm{s}} (t) = 
\sum_{\tilde{m}=1}^{\tilde{M}} \bm{a}^T (\theta,\phi) \bm{e}_{\tilde{m}}^* s_{\tilde{m}} (t) 
	  \label{eq:transmit signal} 
\end{align}
where $\bm{e}_{\tilde{m}}$ is the $\tilde{m}$th column of the transmit beamspace and interpolation matrix $\bm{E}$, $\tilde{\bm{s}} (t) \triangleq [s_1 (t), \ldots, s_{\tilde{M}} (t)]^T$, and $\{ s_{\tilde{m}} (t) \}_{\tilde{m} = 1}^{\tilde{M}}$ is a set of $\tilde{M}$ initial orthogonal waveforms, which can be for example a subset of the set $\{ s_{m} (t) \}_{m = 1}^{M}$ of a larger number of $M$ waveforms in \eqref{eq:radiated signal}.
 
Then the receive array observations can be expressed as
\begin{equation}
\tilde{\bm{x}}(t,q)=\sum_{k=1}^K\beta_k(q)(\tilde{\bm{a}}^T(\theta_k,\phi_k) \tilde{\bm{s}} (t))\bm{b}(\theta_k,\phi_k)+\bm{n}(t,q).
\end{equation}
Thus, after matched-filtering to the ${\tilde{m}}$th waveform, the ${\tilde{m}}$th $N\times 1$ received vector can be expressed as
\begin{align}
\tilde{\bm{y}}_{\tilde{m}}(q) &= \int_T \tilde{\bm{x}}(t,q) s^*_{\tilde{m}} (t)dt \notag \\
&= \sum_{k=1}^K\big(\tilde{\bm{a}}_{\tilde{m}}(\theta_k,\phi_k)\odot\bm{b}(\theta_k,\phi_k)\big) \beta_k(q)+\bm{z}(q)  \notag \\
&= \sum_{k=1}^K\bm f_{\tilde{m}}(\theta_k,\phi_k)\beta_k(q) + \bm{z} (q) \notag \\
&= \bm F_{\tilde{m}}(\theta,\phi)\bm\beta(q) + \bm{z} (q)
\end{align}
where $\bm f_{\tilde{m}}(\theta_k,\phi_k) \triangleq \tilde{\bm{a}}_{\tilde{m}} (\theta_k,\phi_k)\odot\bm{b}(\theta_k,\phi_k)$ and
\begin{equation}
\bm F_{\tilde{m}}(\theta,\phi) \triangleq [\bm f_{\tilde{m}}(\theta_1,\phi_1), \ldots, \bm f_{\tilde{m}}(\theta_K,\phi_K)]
\end{equation}
is the desired steering matrix. Hence, the whole received vector, i.e., the vector obtained by stacking  $\tilde{\bm{y}}_{m} (q)$, $m=1,\ldots,M_1M_2$, one under another, can be expressed as
\begin{equation}
\tilde{\bm{y}} (q) = \bm F(\theta,\phi) \bm{\beta} (q) + \bm{z}(q) \label{eq:TB receive model}
\end{equation}
where $\tilde{\bm{y}} (q) \in \mathbb{C}^{\tilde{M} N\times 1}$, and
\begin{equation}
\bm{F}(\theta,\phi) \triangleq [\bm F_1^T(\theta,\phi),\ldots,\bm F_{\tilde{M}}^T(\theta,\phi)]^T \in\mathbb{C}^{\tilde{M}N\times K} \label{eq:steering matrix} 
\end{equation}
denotes the array steering matrix consisting of individual steering vectors. It is worth noting that \eqref{eq:TB receive model} does not fully capture the multidimensional structure of the data hidden in the 2D TB MIMO radar model when several pulses are considered. Therefore, we develop in the next section a tensor modeling for 2D TB MIMO radar capturing several pulses.

\section{Tensor Formulation Based DOA Estimation}
\subsection{2D TB MIMO Radar Tensor Modeling}
In this section, we show how to fold the 2D TB MIMO radar data  into a tensor form suitable for applying tensor-based ESPRIT method \cite{hass13} for the elevation and azimuth direction finding. Here, we operate directly on the received data using the so-called direct data approach. 

Consider $Q$ pulses, then the received 2D TB MIMO radar data matrix is given as
\begin{equation}
\tilde{\bm{Y}} = \bm F(\theta,\phi) \bm{P} + \bm{Z} \label{eq:MIMO model}
\end{equation}
where $\tilde{\bm{Y}}\in\mathbb{C}^{\tilde{M}N\times Q}$, $\bm{P} \triangleq [\bm{\beta}(1), \bm{\beta}(2), \ldots, \bm{\beta}(Q)]$ contains $K$ vectors of targets' RCS coefficients for $Q$ pulses, and $\bm{Z}$ represents the noise, which is assumed to be Gaussian with zero mean. 

Recalling \eqref{eq:nmode}, the corresponding 2D TB MIMO radar tensor model can be expressed as
\begin{equation}
\bm{\mathcal{Y}} = \bm{\mathcal{A}}\times_R \bm{P} + \bm{\mathcal{Z}} \label{eq:MIMO tensor}
\end{equation}
where steering tensor $\bm{\mathcal{A}}$ is composed by stacking $K$ targets' steering tensor $\bm{\mathcal{A}}_k$ together, $\bm{\mathcal{Z}}$ stands of the noise samples the same as that in \eqref{eq:MIMO model}, and $R$ represents the $R$th mode tensor-matrix product. The relation between the tensor and its matrix unfolding is given by
\begin{equation}
\tilde{\bm{Y}} = \bm{\mathcal{Y}}_{(R)}^T.
\end{equation}

It is worth noting that \eqref{eq:MIMO tensor} represents a general tensor model for the 2D TB MIMO radar with transmit array interpolation. Some practically important special cases, which we will use throughout the paper, are the virtual URA and L-shaped arrays. For the case of virtual L-shaped array, $\bm{\mathcal{Y}}$ becomes a third-order tensor, and its dimensions represent virtual transmit array, receive array, and the temporal pulses, respectively. While for the case of virtual URA, $\bm{\mathcal{Y}}$ becomes a fourth-order tensor with two dimensions carrying elevation and other two dimentions carrying azimuth information. Also, the value of $R$ depends on which type of virtual transmit array we design. For the case of virtual L-shaped array, $R=3$, and for the case of virtual URA, $R=4$.

Thus, we have obtained the tensor form representation of the 2D TB MIMO radar data. This form allows to use some computational efficient methods, e.g., ESPRIT, for elevation and azimuth estimation. Comparing with searching-based DOA estimation methods, such method allows to reduce significantly the computational burden, especially when the model is of large size.

\subsection{Steering Vector and Steering Tensor}
For the case of virtual transmit L-shaped array in the 2D TB MIMO radar model with transmit array interpolation, the received data $\bm{\mathcal{Y}} \in \mathbb{C}^{(M_1 + M_2 - 1)}$ is a third-order tensor. The derivations in this case are similar to that of the case of virtual URA. In addition, the translational invariance property holds for both the virtual transmit L-shaped array and URA in the same way. Therefore, we discuss in details hereafter the case when the transmit array is mapped into a virtual URA, while some details for the case of L-shaped array can be found in \cite{hass13}.

The  relationship between a rank-one third-order tensor and its vector expression is given as follows
\begin{equation}
\bm{a}\odot\bm{b}\odot\bm{c}=\text{vec}(\bm{a}\circ\bm{b}\circ\bm{c}). \label{eq:vectorization}
\end{equation} 
The steering tensor of $k$th target can be expressed as
\begin{equation}
\bm{\mathcal{A}}_k(\theta_k,\phi_k) = \bm{c}(\theta_k)\circ\bm{d}(\phi_k)\circ\bm{b}(\theta_k,\phi_k) \label{eq:tensor piece}
\end{equation}
where $\bm{\mathcal{A}}_k\in\mathbb{C}^{M_1\times M_2\times N}$. Substituting  \eqref{eq:tensor piece} into  \eqref{eq:vectorization}, we obtain the following relationship between the steering vector and the steering tensor
\begin{equation}
\text{vec}(\bm{\mathcal{A}}_k(\theta_k,\phi_k))=\bm{c}(\theta_k)\odot\bm{d}(\phi_k)\odot\bm{b}(\theta_k,\phi_k) . \label{eq:outer}
\end{equation}
It is clear that the array steering tensor obeys the translational invariance property with respect to $\bm{c}(\theta_k)$ or $\bm{d}(\phi_k)$, which makes tensor-based ESPRIT methods applicable for this scheme. 

\subsection{Tensor-Based Signal Subspace}
Consider the case of virtual transmit URA as the most general example. In this case, the received tensor has four modes, $R=4$. The main motivation behind expressing the 2D TB MIMO radar model in the tensor form is that it allows for a more accurate elevation and azimuth estimation through tensor decomposition. The HOSVD of the tensor \eqref{eq:MIMO tensor} is given by
\begin{equation}
\bm{\mathcal{Y}}=\bm{\mathcal{S}}\times_1 \hat{\bm{U}}_1\times_2\hat{\bm{U}}_2\times_3\hat{\bm{U}}_3\times_4\hat{\bm{U}}_4 \label{eq:HOSVD}
\end{equation}
where $\bm{\mathcal{S}}\in\mathbb{C}^{M_1\times M_2\times N\times Q}$ represents the core tensor, and $\hat{\bm{U}}_r, r=1,\ldots,4$ are unitary matrices. The tensor-based signal subspace can be then obtained by truncating noisy observation tensor $\bm{\mathcal{Y}}$ as
\begin{equation}
\hat{\bm{\mathcal{U}}}^{[s]}=\bm{\hat{\mathcal{S}}}^{[s]}\times_1\hat{\bm{U}}_1^{[s]}\times_2\hat{\bm{U}}_2^{[s]}\times_3\hat{\bm{U}}_3^{[s]} \label{eq:sb}
\end{equation}
where $\bm{\hat{\mathcal{S}}}^{[s]}\in\mathbb{C}^{K\times K}$ and $\hat{\bm{U}}_r^{[s]} \in \mathbb{C}^{M_r\times K}$ are the core tensor and $r$th mode signal subspace estimate, respectively. Then we can give the relationship between the signal subspace of noisy observation matrix $\tilde{\bm{Y}}$ and the matrix form of the HOSVD-based subspace. The SVD of $\tilde{\bm{Y}}$ can be expressed as 
\begin{equation} \label{eq:svd}
	\tilde{\bm{Y}}=
	\begin{bmatrix} \hat{\bm{U}}_s &  \hat{\bm{U}}_n\end{bmatrix} 
	\begin{bmatrix} \hat{\bm{\varSigma}}_s & \bm{0} \\ 
	\bm{0} & \hat{\bm{\varSigma}}_n \end{bmatrix} 
	\begin{bmatrix} \hat{\bm{V}}_s &  \hat{\bm{V}}_n \end{bmatrix}^H
\end{equation}
where $\hat{\bm{U}}_s$ and $\hat{\bm{U}}_n$ are the matrices of the signal and noise subspaces estimates, respectively, while $\hat{\bm{\varSigma}}_s$ and $\hat{\bm{\varSigma}}_n$ are the corresponding diagonal matrices of singular values. Indeed, HOSVD is actually performing SVD along different unfoldings of a tensor. Thus, the SVD of the $r$th mode unfolding of the data tensor $\bm{\mathcal{Y}}$ can be expressed as
\begin{equation} \label{eq:hosvdr}
\bm{\mathcal{Y}}_{(r)}=
\begin{bmatrix}\hat{\bm{U}}_r^{[s]} &  \hat{\bm{U}}_r^{[n]}\end{bmatrix} 
\begin{bmatrix}\hat{\bm{\varSigma}}_r^{[s]} & \bm{0} \\ 
\bm{0} & \hat{\bm{\varSigma}}_r^{[n]} \end{bmatrix} 
\begin{bmatrix}\hat{\bm{V}}_r^{[s]} & \hat{\bm{V}}_r^{[n]}	\end{bmatrix}^H
\end{equation}
where $\hat{\bm{U}}_r^{[s]}$, $\hat{\bm{U}}_r^{[n]}$, $\hat{\bm{\varSigma}}_r^{[s]}$, $\hat{\bm{\varSigma}}_r^{[n]}$ are the matrices composed of the signal and noise eigenvectors, and signal and noise eigenvalues, respectively, of the $r$th mode unfolding of $\bm{\mathcal{Y}}$. We have called  $\hat{\bm{U}}_r^{[s]}$ previously as $r$th mode signal subspace estimate.

There is the following relationship between the steering tensor and the signal subspace
\begin{equation}
\bm{\mathcal{A}}\approx\hat{\bm{\mathcal{U}}}^{[s]}\times_{4}\bm{T}
\end{equation}
where $\bm{T}$ denotes a nonsingular transform matrix.

The rotational invariance property holds for the tensor-based signal subspace, and it can be exploited for estimating both elevation and azimuth. If there are no interpolation errors in the mapping of the actual transmit array to a virtual one, we have
\begin{align}
\hat{\bm{\mathcal{U}}}^{[s]}\times_1\bm{J}_1^{(1)}\times_{4}\bm{\varTheta}_r \approx \hat{\bm{\mathcal{U}}}^{[s]}\times_1\bm{J}_2^{(1)} \label{eq:RIP1}\\
\hat{\bm{\mathcal{U}}}^{[s]}\times_2\bm{J}_1^{(2)}\times_{4}\bm{\varPhi}_r \approx \hat{\bm{\mathcal{U}}}^{[s]}\times_2\bm{J}_2^{(2)} \label{eq:RIP2}
\end{align}
where $\bm{J}_1^{(r)}$ and $\bm{J}_2^{(r)}$ stand for the selection matrices of each virtual array, and are given as 
\begin{align}
\bm{J}_1^{(r)} \triangleq \big[\bm{I}_{M_r} \quad \bm{0}_{M_r\times 1}\big]\\
\bm{J}_2^{(r)} \triangleq \big[\bm{0}_{M_r\times 1} \quad \bm{I}_{M_r}\big]
\end{align}
for $r=1,2$. In \eqref{eq:RIP1} and \eqref{eq:RIP2}, the rotational invariance property is satisfied in between the left and right-hand sides of the expressions. 

Performing eigenvalue decomposition (EVD) of $\bm{\varTheta}_r$ and $\bm{\varPhi}_r$, we obtain
\begin{align}
	\bm{\varTheta}_r=\bm{T}_{1r}\bm{\varTheta}\bm{T}^{-1}_{1r}\\
	\bm{\varPhi}_r=\bm{T}_{2r}\bm{\varPhi}\bm{T}^{-1}_{2r}
\end{align}
where $\bm{T}_{1r}$ and $\bm{T}_{2r}$ are nonsingular (invertible) matrices since the targets are independent, $\bm{\varTheta}$ and $\bm{\varPhi}$ are the diagonal matrices carrying the elevation and azimuth, respectively, and given as
\begin{align}
\bm{\varTheta} \triangleq \text{diag}\{ [e^{-j\pi\sin\theta_1\cos\phi_1}, \ldots, e^{-j\pi\sin\theta_K\cos\phi_K}]\} \\
\bm{\varPhi} \triangleq \text{diag}\{ [e^{-j\pi\sin\theta_1\sin\phi_1}, \ldots, e^{-j\pi\sin\theta_K\sin\phi_K}]\}.
\end{align}

\subsection{HOSVD-Based DOA Estimation}
The above developed 2D TB MIMO radar model has a similar form with the so-called  $R$-dimensional ($R$-D) harmonic retrieval (HR) problem \cite{Haardt08}, \cite{Sun13}. The main difference is that the proposed model does not necessarily obey the Vandermonde structure at the receive array. However, it is not needed, since we can still solve the DOA estimation problem through taking advantage of the translational invariance property enforced at the transmit array while neglecting the structure of the receive array. If we exploit the whole signal subspace, the problems of finding $\bm{\varTheta}_r$ and $\bm{\varPhi}_r$ can be expressed in the form of the following two tensor-based LS problems
\begin{align}
\bm{\varTheta_r}=\arg\min\limits_{\bm{\varTheta_r}}\parallel \bm{\mathcal{U}}^{[s]}\times_1 \bm{J}_1^{(1)} \times_4 \bm \varTheta_r- \bm{\mathcal{U}}^{[s]}\times_1 \bm{J}_2^{(1)} \parallel_\text{H} \\
\bm{\varPhi_r}=\arg\min\limits_{\bm{\Phi_r}}\parallel \bm{\mathcal{U}}^{[s]}\times_2 \bm{J}_1^{(2)} \times_4 \bm \varPhi_r - \bm{ \mathcal{U}}^{[s]}\times_2 \bm{J}_2^{(2)} \parallel_\text{H} . 
\end{align}

These two tensor-based LS problems are equivalent to  
\begin{align} 
\bm{\varTheta}_r=\text{arg}\min_{\bm{\varTheta}_r} \| \bm{\varTheta}_r \bm{\mathcal{U}}^{[s]}_{(4)} \bm{\varOmega}_{11}^T - \bm{\mathcal{U}}^{[s]}_{(4)}\bm{\varOmega}_{21}^T \|_{\rm F} \label{eq:fse}\\
\bm{\varPhi}_r=\text{arg}\min_{\bm{\varPhi}_r} \| \bm{\varPhi}_r \bm{\mathcal{U}}^{[s]}_{(4)} \bm{\varOmega}_{12}^T -\bm{\mathcal{U}}^{[s]}_{(4)}\bm{\varOmega}_{22}^T \|_{\rm F} \label{eq:fsa} 
\end{align}
where 
\begin{align}
	\bm{\varOmega}_{ir} &\triangleq \bm{I}_{L_1}^{(r)} \otimes \bm{J}_i^{(r)} \otimes \bm{I}_{L_2}^{(r)} \label{eq:omega}  \\
	L_1^{(r)} &\triangleq \prod_{i=1}^{r-1}M_i, \quad L_2^{(r)} \triangleq \prod_{i=r+1}^RM_i
\end{align}
for $i=1,2$ and $p=1,2$. Solving \eqref{eq:fse} and \eqref{eq:fsa}, yields,   
\begin{align}
	\bm{\varTheta}^T=(\bm{\varOmega}_{11}[\bm{\mathcal{U}}^{[s]}_{(4)}]^T)^\dagger\bm{\varOmega}_{21}[\bm{\mathcal{U}}^{[s]}_{(4)}]^T  \label{eq:hosvdele}\\
	\bm{\varPhi}^T=(\bm{\varOmega}_{12}[\bm{\mathcal{U}}^{[s]}_{(4)}]^T)^\dagger\bm{\varOmega}_{22}[\bm{\mathcal{U}}^{[s]}_{(4)}]^T. \label{eq:hosvdazi}
\end{align}

It has to be noticed however that since this method exploits the whole signal subspace when estimating elevation and azimuth, the DOA estimation errors can be enlarged if there exist interpolation errors in mapping the actual transmit array into a virtual one, especially in the high SNR region. 

\subsection{TEV-Based DOA Estimation}
Another tensor decomposition method proposed in \cite{Sun13} and called as tensor approach based on eigenvectors (TEV) can be used as well for DOA estimation in our framework as an alternative to \eqref{eq:hosvdele}--\eqref{eq:hosvdazi}. In accordance with the TEV-based method, we divide the tensor-based signal subspace into several small parts along the targets' dimension as
\begin{equation}
\hat{\bm{\mathcal{U}}}^{[s]}=\bm{\mathcal{Q}}_1\sqcup_{4}\bm{\mathcal{Q}}_2\sqcup_{4}\cdots\sqcup_{4}\bm{\mathcal{Q}}_K \label{eq:subtensor}.
\end{equation}
For $k$th signal dimension, we have 
\begin{equation}
	\bm{\mathcal{A}}_k=\bm{\mathcal{Q}}_k\times\bm{T}_k， \quad k=1,2,\ldots,K \label{eq:subk}
\end{equation}
where $\bm{T}_k$ is a non-singular matrix. Recalling \eqref{eq:tensor piece}, we find out that $\bm{\mathcal{Q}}_k$ represents the signal subspace associated with $k$th signal dimension. Applying HOSVD to $\bm{\mathcal{Q}}_k$, yields,
\begin{align}
	\bm{\mathcal{Q}}_k=\xi_{k}\times_1\bm{q}_{k,1}\times_2\bm{q}_{k,2}\times_3\bm{q}_{k,3} . \label{eq:tevmoder}
\end{align}
It can be seen that $\bm{q}_{k,r}$ spans the same subspace as the $r$th dimension of $\bm{\mathcal{A}}_k$.  Using the linear prediction property \cite{So10}, we can apply ESPRIT method to $\bm{q}_{k,1}$ and $\bm{q}_{k,2}$ when estimating $\mu_k$ and $\nu_k$ as
\begin{align}
\bm{J}_1^{(1)}\bm{q}_{k,1} = e^{-j\pi\mu_k} \bm{J}_2^{(1)}\bm{q}_{k,1} \\
\bm{J}_1^{(1)}\bm{q}_{k,2} = e^{-j\pi\nu_k} \bm{J}_2^{(1)}\bm{q}_{k,2}.	
\end{align}

The TEV-based method operates on each dimension of the signal subspace separately, which makes the proposed method more robust to interpolation errors because of the mapping of the actual transmit array into a virtual one. In addition, the auto-pairing has to be performed. It is technical, but otherwise straightforward, and it has been addressed, for example, in \cite{haardt98} as well as in a number of other papers.

\section{CRB and Influence of Interpolation Errors on DOA Estimation}
\subsection{CRB}
A useful statistical bound for evaluating the limiting DOA estimation performance in our framework is the deterministic CRB \cite{Stoica89}. It can be derived for the considered 2D TB MIMO radar model with transmit array interpolation, i.e., model \eqref{eq:TB receive model}, in close-form. It is worth mentioning that the proposed 2D TB MIMO radar model with transmit array interpolation differs from the MIMO radar model given in \eqref{eq:receive model} due to the fact that the interpolated transmit array is used in \eqref{eq:TB receive model}, while the original transmit array is used in \eqref{eq:receive model}. Thus, the CRB for the model \eqref{eq:TB receive model} will be a function of the transmit interpolation matrix, and hence, the interpolation error will also be considered. 
	
It is straightforward to see that the model \eqref{eq:TB receive model} can be regarded as a special type of the deterministic $R$-D HR model as we discussed in Subsection~IV.D. The corresponding closed-form CRB expression for the deterministic $R$-D HR model has been derived in \cite{Liu06}. In our case, the unknown parameter vector of interest is
\begin{equation} \label{parametersCRB}
\bm{\omega} \triangleq [\bm{\theta}^T, \, \bm{\phi}^T]^T
\end{equation} 
and the CRB for accuracy of the parameter vector \eqref{parametersCRB} estimate is therefore can be written in our notations as
\begin{align}
\text{CRB} (\bm{\omega}) = \frac{\sigma_n^2}{2} \text{diag} \bigg \{&\sum_{q=1}^{Q} \text{Re} \big\{\bm{C}_q^H \bm{D}^H \notag\\&
\times (\bm{I} - \bm{F} (\bm{F}^H \bm{F})^{-1}\bm{F}^H) \bm{D} \bm{C}_q  \big \}\bigg \}^{-1}
\end{align}
where $\sigma_n^2$ is the noise variance and
\begin{align}
\bm{C}_q \triangleq& \bm{I}_2 \otimes \text{diag} (\bm \beta(q)) \\
\bm{D} \triangleq& [\bm{D}_{\rm e}, \bm{D}_{\rm a}] \\
\bm{D}_{\rm e} \triangleq& \left(\bm{E}^H \left( \bm{W}_{\rm te} \ast \bm{A} \right) \right) \odot \bm{B} + \bm{E}^H \bm{A} \odot (\bm{W}_{\rm re} \ast \bm{B}) \\
\bm{D}_{\rm a} \triangleq& \left( \bm{E}^H \left( \bm{W}_{\rm ta} \ast \bm{A} \right) \right) \odot \bm{B} + \bm{E}^H \bm{A} \odot(\bm{W}_{\rm ra} \ast \bm{B}) 
\end{align}
with $\bm{W}_{\rm te}, \bm{W}_{\rm ta}, \bm{W}_{\rm re}, \bm{W}_{\rm ra}$ being the first order derivation of the steering matrix of the transmit array and receive array with respect to elevation and azimuth, respectively. Here we also use notations $\bm{D}$, $\bm{D}_{\rm e}$, $\bm{D}_{\rm a}$, $\bm{A}$, $\bm{F}$, $\bm{B}$ instead of, respectively, the full notations $\bm{D} (\theta,\phi)$, $\bm{D}_{\rm e}  (\theta,\phi)$, $\bm{D}_{\rm a}  (\theta,\phi)$, $\bm{A}  (\theta,\phi)$, $\bm{F}(\theta,\phi)$, $\bm{B}  (\theta,\phi)$ for brevity.

\subsection{DOA Estimation Bias} \label{sec:bias}
Obviously, the interpolation of the actual transmit array into a virtual one within a sector of interest by solving, for example, the minimax optimization problem \eqref{eq:OptObj_2}--\eqref{eq:OptConst_2} is not perfect. Here we investigate the influence of the interpolation errors to the DOA estimation accuracy.  

The interpolation errors can be expressed as 
\begin{equation}
\Delta\bm{A}(\theta,\phi) \triangleq {\bm{E}}^H {\bm{A}} (\theta,\phi) - \tilde{\bm{A}} (\theta,\phi). \label{eq:interpolationerror}
\end{equation}
where $\Delta\bm{A}(\theta,\phi)$ is regarded as model errors and can be calculated once the transmit array interpolation matrix is designed. As a result, the interpolation errors will cause a bias in DOA estimations through a perturbation of the signal subspace. The DOA estimation performance degradation caused by imperfect interpolation can even dominate over the finite samples effect in high SNR region \cite{Hyberg04}, \cite{Hyberg05}. 

Let us thus analyse the effect of the interpolation errors on DOA estimation from the subspace perturbation point of view which is suitable for the direct data approaches. For simplicity, we isolate model errors from finite sample effects by neglecting the observation noise, i.e., the SNR is assumed to be large enough. The SVD of the noiseless observation model is then given as
\begin{equation} \label{eq:noiselesssvd}
\bar{\bm{Y}}_0=
\begin{bmatrix}\bm{U}_s &  \bm{U}_n\end{bmatrix} 
\begin{bmatrix}\bm{\varSigma}_s & \bm{0} \\ \bm{0} & \bm{0} \end{bmatrix} 
\begin{bmatrix}\bm{V}_s &  \bm{V}_n	\end{bmatrix}^H
\end{equation}
where $\bm{U}_s$ and $\bm{V}_s$ represent, correspondingly, the left and right singular vectors associated with $K$ non-zero singular values, which form the diagonal of the diagonal submatrix $\bm{\Sigma}_s$, while $\bm{U}_n$ and $\bm{V}_n$ contain, respectively, the left and right singular vectors associated with the zero singular values.

We note that the matrix-based signal subspace is equivalent to the tensor-based signal subspace in the absence of noise \cite{Haardt08}. Thus, we can use just the matrix-based signal subspace hereafter. Here we define the perturbations of direction cosine functions as 
\begin{align}
	&\Delta\mu_k \triangleq \hat{\mu}_k-\mu_k, \quad k = 1 ,2, \ldots, K \\
	&\Delta\nu_k \triangleq \hat{\nu}_k-\nu_k, \quad k = 1 ,2, \ldots, K
\end{align} 
where $\hat{\mu}_k$ and $\hat{\nu}_k$ denote $k$th estimated direction cosine functions, while $\mu_k$ and $\nu_k$ denote the $k$th true direction cosine functions without the effect due to the interpolation errors.

In \cite{Li92}, the perturbed DOA estimation performance analysis has been performed for the case of virtual ULA and 1D matrix-based ESPRIT method. Our case of the 2D TB MIMO radar model can be regarded as a multi-dimension extension of the model considered in \cite{roemer14}.  Otherwise, such extension is straightforward, and using the results in \cite{Li92}, the expectation of mean-squared error (MSE) with the perturbations of direction cosine functions $\Delta\mu_k$ and $\Delta\nu_k$ for the proposed 2D TB MIMO radar model with array interpolation can be found as  
\begin{align}
\mathbb{E}_{\Delta\bm{A}(\theta,\phi)} \{ (\Delta \mu_k)^2 \} = \frac{\sigma_{\rm app}^2}{2\pi^2} \| \bm{\alpha}_{k1} \|^2_2 \cdot \| \bm{\beta}_{k1} \|^2_2, \quad k=1,\ldots,K \label{eq:permu} \\
\mathbb{E}_{\Delta\bm{A}(\theta,\phi)} \{ (\Delta \nu_k)^2 \} = \frac{\sigma_{\rm app}^2}{2\pi^2} \| \bm{\alpha}_{k2} \|^2_2 \cdot \| \bm{\beta}_{k2} \|^2_2, \quad k=1,\ldots,K \label{eq:pernu}
\end{align}
where $\bm{\alpha}_{kr} = \bm{p}_{kr}^T (\bm{\varOmega}_{1r} \bm{U}_s)^\dagger (\bm{\varOmega}_{2r} / \lambda_{kr} - \bm{\varOmega}_{1r}) \bm{U}_n \bm{U}_n^H$, $\bm{\beta}_{kr} = \tilde{\bm{A}} (\theta, \phi)^\ddagger \bm{U}_s \bm{q}_{kr}$, $\bm{p}_{kr}$ and $\bm{q}_{kr}$ are the left and right orthgonal eigenvectors of $\bm{\varTheta}$ and $\bm{\varPhi}$ associated with $k$th eigenvalue $\lambda_k$, respectively, and $r=1,2$. The expressions for $\bm{\varOmega}_{ir}, i=1,2$ have been given in \eqref{eq:omega}, and $\sigma_{\rm app}^2$ denotes the array interpolation (approximation) errors, which can be formally expressed at a point $(\theta, \phi)$ as 
\begin{equation}
\sigma_{\rm app}^2 = \frac{1}{M_1 M_2} \| \Delta\bm{A}(\theta,\phi) \|_2^2. \label{eq:errors}
\end{equation}
Note that $\sigma_{\rm app}^2$ is different from the interpolation error tolerance $\Delta$ as we defined it in \eqref{eq:OptConst_2}. In particular, $\sigma_{\rm app}^2$ represents the exact array interpolation error at $(\theta, \phi)$, while the interpolation error tolerance $\Delta$  represents the worst array interpolation condition. 

Based on \eqref{eq:permu}, \eqref{eq:pernu}, and \eqref{eq:errors}, we can finally derive the expectation of MSEs of the elevation and azimuth estimation biases. The details of the derivations are given in Appendix A, and the resulting  $\mathbb{E} \{ (\Delta\theta_k)^2 \}_{\Delta\bm{A}(\theta,\phi)}$ and $\mathbb{E} \{ (\Delta\phi_k)^2 \}_{\Delta\bm{A}(\theta,\phi)}$ are given as
\begin{align}
	\mathbb{E}_{\Delta\bm{A}(\theta,\phi)} \{ (\Delta\theta_k)^2 \} &= \frac{\mu_k^2}{\mu_k^2 + \nu_k^2} \mathbb{E}_{\Delta\bm{A}(\theta,\phi)} \{ (\Delta\mu_k)^2 \} \nonumber \\ 
	& + \frac{\nu_k^2}{\mu_k^2+\nu_k^2} \mathbb{E}_{\Delta\bm{A}(\theta,\phi)} \{ (\Delta\nu_k)^2 \} \label{eq:ele bias} \\
	\mathbb{E}_{\Delta\bm{A}(\theta,\phi)} \{ (\Delta\phi_k)^2 \} &= \frac{\nu_k^2}{(\mu_k^2+\nu_k^2)^2} \mathbb{E}_{\Delta\bm{A}(\theta,\phi)} \{ (\Delta\mu_k)^2 \} \nonumber \\
	& + \frac{\mu_k^2}{(\mu_k^2+\nu_k^2)^2} \mathbb{E}_{\Delta\bm{A}(\theta,\phi)} \{ (\Delta\nu_k)^2 \}. \label{eq:azi bias}
\end{align}
In addition, the relationships between the perturbation of direction cosine functions and DOA estimation biases are also given in Appendix B. 

\subsection{Look-Up Table}
As shown by \eqref{eq:ele bias} and \eqref{eq:azi bias}, the interpolation errors cause undesirable DOA estimation bias. However, we fortunately can compensate the bias to some extent through building an off-line look-up table over the spatial sector of interest. The look-up table building procedure is summarized in Table \ref{table:lookup}. Then this table can be used to map the DOAs estimated from noisy model to more precise values in order to obtain the corresponding accurate estimates. 

\begin{table}[!t] 
\renewcommand{\arraystretch}{1.6}
\caption{Algorithm for Building Offline DOA Look-up Table}
\label{table_example} \label{table:lookup}
\centering
\begin{tabular}{m{8cm}}  
\toprule[.8pt] 
\hline
(\romannumeral1) Consider the transmit array interpolation model \eqref{eq:TB receive model} and assume a single source located at $(\theta, \phi)\in(\Theta,\Phi)$; map ($\theta, \phi$) to ($\mu, \nu$) using $\mu=\sin\theta\cos\phi$ and $\nu=\sin\theta\sin\phi$. \\
(\romannumeral2) Assuming that the observation model is noise free, build the corresponding noise-free tensor model of \eqref{eq:MIMO tensor}. \\
(\romannumeral3) Apply the proposed DOA estimation methods to the noise-free tensor model  to obtain  ($\hat\mu,  \hat\nu$). \\
(\romannumeral4) Repeat (\romannumeral1)--(\romannumeral3) over a fine 2D grid within the 2D spatial sector of interest. \\
(\romannumeral5) Form two look-up tables mapping $\mu$ to $\hat\mu$ and $\nu$ to $\hat\nu$. \\
\bottomrule[.8pt]
\end{tabular}
\end{table}

\section{Simulation Results}
In this section, we evaluate the effectiveness of the proposed method in terms of simulations. Since the TB and interpolation matrix design can be done off-line, we compare computational burden of the matrix-based and tensor-based DOA estimation methods with virtual transmit URA. It is worth noting that the computational complexity for the $R$-D HR approach has been calculated in \cite{Haardt08} and \cite{Sun13}. The major computational complexities of the DOA estimation methods are the SVD and HOSVD parts, which require the complexities of $k_t K \prod_{i=1}^2 M_iQ$ and $4 k_t K\prod_{i=1}^ 2M_iQ$ for direct data approach, respectively. Here $k_t$ is a constant that depends on a specific method used for performing SVD \cite{Golub96}. Therefore, the tensor-based DOA estimation method has a higher, by a factor of 4, computational burden than the matrix-based one. Also, the computational burden of the DOA estimation method used in the case of mapping to virtual transmit L-shaped array is smaller than that of the case of mapping to virtual transmit URA.  

In all examples, the mono-static MIMO radar with 64 co-located transmit antenna elements and 8 receive antenna elements is assumed. The transmit array is an irregular planar array for which each transmit antenna element is placed near the regular grid of $8 \times 8$ URA of size of $4\lambda$ for each column and row. To realize this structure, we firstly generate an URA of size $8\times 8$ of length $4\lambda$ for each column and row, and then add random displacements, which obey uniform distributions $[-\frac{\lambda}{4},\frac{\lambda}{4}]$, to the antenna elements' positions.  The receive array consists of the antenna elements which are randomly selected from the transmit array. The transmit interpolation technique expressed in terms of the optimization problem \eqref{eq:OptObj_2}--\eqref{eq:OptConst_2} is used to obtain virtual transmit array. In \eqref{eq:OptObj_2}--\eqref{eq:OptConst_2}, the $l_\infty$-norm is used in the objective function and $l_1$-norm is used in the constraints. The problem \eqref{eq:OptObj_2}--\eqref{eq:OptConst_2} minimizes sidelobes on a grid outside of the sector of interest, while keeping satisfactory accuracy within the sector of interest for each point on a grid. Thus, the use of $l_\infty$-norm in the minimax objective function \eqref{eq:OptObj_2}	further guarantees that the worst (highest) sidelobe level is minimized, while the use of $l_1$-norm in the constraints \eqref{eq:OptConst_2} provides the least physical displacement of the actual antenna elements when they are mapped to the virtual elements. It is just because for each point on a grid within the sector of interest the sparsest solution, i.e., the solution that requires the smallest physical displacement between the actual and virtual antenna elements, is preferred by using $l_1$-norm.

\subsection{TB and Interpolation Matrix}
In all examples of this subsection, the targets are assumed to be located within a known spatial sector. The spatial sector of interest is set to be $\Theta = [30^\circ, 40^\circ]$ and $\Phi = [65^\circ, 75^\circ]$. We also design a transition zone at each side of the elevation and azimuth domains of $20^\circ$ and $15^\circ$, respectively. 
\begin{figure}[tbp]
	\centering
	\vspace{0.5em}
	\flushleft
	\subfigure[Interpolation errors tolerance $\Delta=0.1$]{
		\begin{minipage}[b]{0.4\textwidth}
			\includegraphics[height=3.8in,width=5.2in]{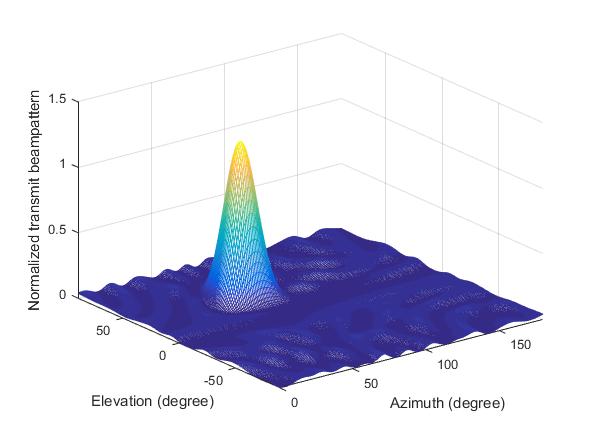}
			
		\end{minipage}
	} \\
	\subfigure[Interpolation errors tolerance $\Delta=0.01$]{
		\begin{minipage}[b]{0.4\textwidth}
			\includegraphics[height=3.8in,width=5.2in]{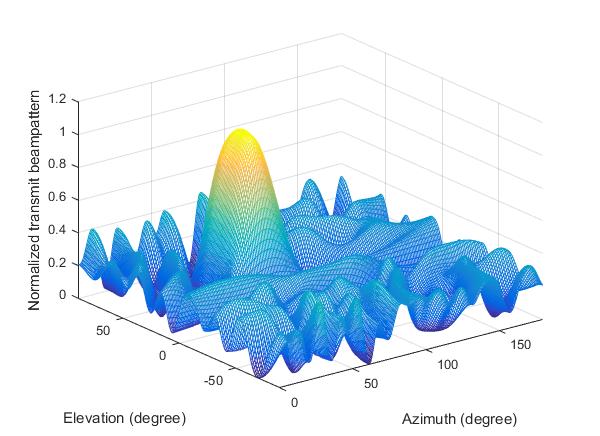}
			
		\end{minipage}
	}\caption{Mono-static MIMO radar transmit beampattern with virtual URA.}
	\label{fig:URA}
\end{figure}

\begin{figure}[htbp]
	\centering
	\vspace{0.5em}
	\flushleft
	\subfigure[Interpolation errors tolerance $\Delta=0.1$]{
		\begin{minipage}[b]{0.4\textwidth}
			\includegraphics[height=3.8in,width=5.2in]{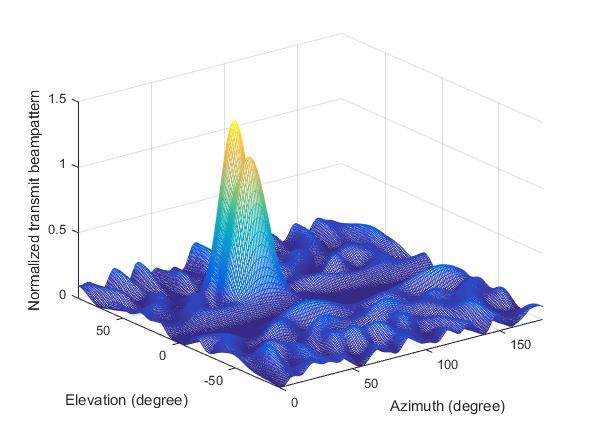}
			
		\end{minipage}
	} \\
	\subfigure[Interpolation errors tolerance $\Delta=0.01$]{
		\begin{minipage}[b]{0.4\textwidth}
			\includegraphics[height=3.8in,width=5.2in]{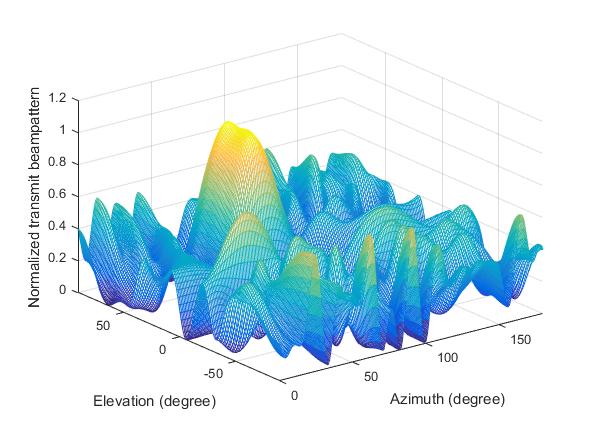}
			
		\end{minipage}
	}\caption{Mono-static MIMO radar transmit beampattern with virtual L-shaped array.}
	\label{fig:L}
\end{figure}

First, we show the transmit beampatterns obtained for the 2D TB MIMO radar designs with interpolation, and then investigate the interpolation errors within the desired spatial sector. 

In Fig. \ref{fig:URA}, the resulting transmit beampatterns obtained for the 2D TB MIMO radar with interpolation are shown. The actual transmit array is mapped here to the virtual URA with 4 antenna elements in each row and column, and the distance between adjacent antenna elements in the virtual URA is set to half wavelength. The interpolation errors are $0.1$ and $0.01$ in Fig. \ref{fig:URA}(a) and Fig. \ref{fig:URA}(b), respectively. It can be observed that the sidelobes of Fig. \ref{fig:URA}(a) are much smaller than the sidelobes in the Fig. \ref{fig:URA}(b), while the main beam of Fig. \ref{fig:URA}(a) is sharper than that in Fig. \ref{fig:URA}(a). Both figures show that the transmit beam powers are distributed mainly and almost uniformly within the spatial sector of interest. The TB with smaller interpolation errors focus more power with the spatial sector of interest. 

In one more example, we investigate the transmit beampattern when the actual transmit array is mapped to the desired transmit array with L-shaped and 4 virtual antenna elements in the row and the column of such virtual array. The interpolation errors are set as before to $0.1$ and $0.01$. Fig. \ref{fig:L} shows the corresponding beampatterns. It can be observed from the figure that by TB design we can suppress the sidelobes and focus the transmit power within the spatial sector of interest. However, the transmit power is not uniformly distributed over the spatial sector. This may lead to DOA estimation performance degradation as compared to the previously tested example where the actual array has been mapped to the virtual URA of a larger number of virtual elements. For smaller interpolation tolerance, the sidelobes become larger because of the tradeoff between the interpolation error and transmit beamforming performance. Indeed, a sufficiently high accuracy of transmit interpolation has to be guaranteed within the sector of interest. When interpolation errors are smaller within the sector of interest, the accuracy of transmit array interpolation is higher, but there are less degrees of freedom left for sidelobes control. Therefore, sidelobes go higher as it can be seen by comparing to each other subfigures (a) and (b) in Figs.~\ref{fig:URA} and \ref{fig:L}.

\begin{figure}[tbp]
	\centering
	\includegraphics[height=3.8in,width=5.2in]{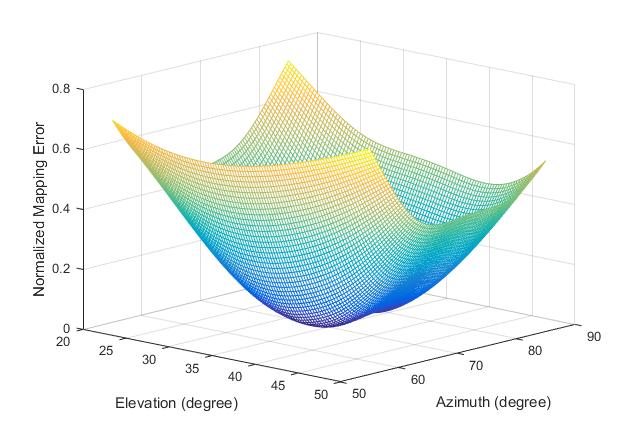}
	\caption{Normalized interpolation errors versus elevation and azimuth.}
	\label{fig:error01}
\end{figure}

In the last example of this subsection, we study the performance of interpolation errors versus elevation and azimuth within the spatial sector of interest. For this, we express the normalized interpolation error with respect to elevation and azimuth at $(\theta,\phi)$ as 
\begin{equation} \label{normilizederror}
\epsilon(\theta,\phi)=\frac{ \| \bm{E}^H \bm{a}(\theta,\phi) - \tilde{\bm{a}} (\theta,\phi) \|_2}{\| \tilde{\bm{a}}(\theta,\phi) \|_2}.
\end{equation}
The transmit array with 64 colocated antenna elements is interpolated to a virtual URA of size $4\times 4$. The array interpolation tolerance is set as $\Delta=0.1$. Fig.~\ref{fig:error01} shows the graph for the normalized interpolation errors \eqref{normilizederror} as a function of elevation and azimuth. It can be seen from the figure that the interpolation error is constrained below a certain level, and it is distributed almost uniformly within the spatial sector of interest. 

\subsection{DOA Estimation Performance}
We evaluate the DOA estimation performance in terms of the root MSE (RMSE), which is computed, for elevation as
\begin{equation}
\mathrm{RMSE}_{\theta} \triangleq \sqrt{\mathbb{E}\Bigg\{\frac{1}{K} \sum_{k=1}^K (\hat{\theta}_k - \theta_k )^2 \Bigg\}} \label{eq:rmse}.
\end{equation}
The RMSE expression for azimuth is the same as \eqref{eq:rmse} where elevation is replaced with azimuth parameter. SNR is defined as $10\text{log}_{10}( \|\tilde{\bm{A}} \bm{B} \|_{\rm F}^2/ \| \bm{Z} \|_{\rm F}^2)$. Performance of the matrix-based and tensor-based ESPRIT and 2D spectral MUSIC methods for elevation and azimuth estimation comparing with deterministic CRB as well as probability of resolving two targets are investigated to examine the effectiveness of the proposed methods.

For all simulation examples in this subsection, the number of Monte Carlo trials for each experiment is set as 1000. Note that the number of pulses is smaller than the size of the 2D TB MIMO radar.  

\begin{figure}[tbp]
	\centering
	\includegraphics[height=3.8in,width=5.2in]{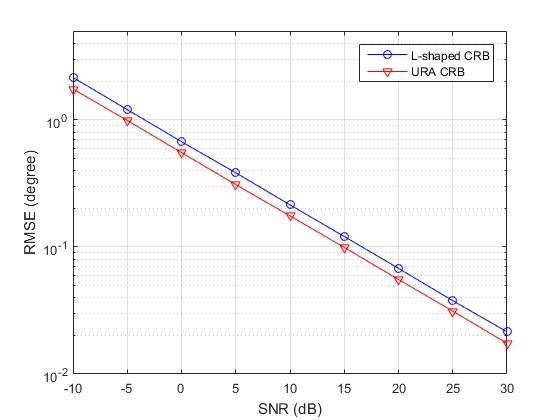}
	\caption{CRB comparison of virtual URA and L-shaped array with interpolation error  $\Delta=0.1$, $Q$=8.}
	\label{fig:CRB01q4}
\end{figure}

\begin{figure}[tbp]
	\centering
	\includegraphics[height=3.8in,width=5.2in]{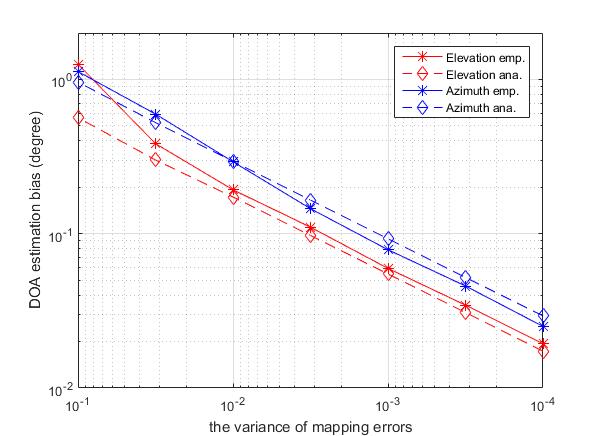}
	\caption{Elevation and azimuth estimation bias versus interpolation/mapping error, $4\times 4$ elements virtual transmit URA, 8 elements receive array,  $Q$=8, $K=2$.  }
	\label{fig:bias}
\end{figure}

In the first example, we plot the CRB associated with the L-shaped array and URA to testify the performance of the two desired virtual transmit array structures. We set the interpolation error tolerance to $0.1$ for both virtual transmit array structures tested. The virtual L-shaped array has 4 elements along X-axis and 4 elements along Y-axis, while the size of the virtual URA is $4\times 4$.  The spatial sector of interest is set as $\Theta=[30^\circ,40^\circ]$ and $\Phi=[65^\circ,75^\circ]$.  It can be observed from  Fig.~\ref{fig:bias} that the deterministic CRB on the DOA estimation for the case of virtual transmit URA is lower than that of the virtual L-shaped transmit array.

In the second example, we examine the influence of the interpolation errors $\sigma^2_a$ on the DOA estimation accuracy. Indeed, the matrix- and tensor-based ESPRIT algorithms are sensitive to interpolation errors since the interpolated transmit array satisfies the translational invariance property only approximatly. Here, the transmit array with 64 colocated antenna elements is interpolated to a $4\times 4$ virtual URA. For given $\Delta$, we obtain interpolation matrix through transmit array interpolation design, thus, the interpolation error $\sigma^2_a$ can be calculated using \eqref{eq:errors}. The empirical DOA estimation error is compared to the one obtained based on the analytical expressions derived in Subsection~\ref{sec:bias}, and the results are shown in Fig.~\ref{fig:bias}. The figure verifies that the empirical DOA estimation bias agrees with the analytically computed one. For example, if $\Delta=0.1$ in the TB and interpolation matrix design in \eqref{eq:interpolation}, the corresponding interpolation error is $\sigma_{\rm app}^2 = 5 \times 10^{-3}$. As a result, we have $\Delta \theta = 0.1^\circ$ and $\Delta \phi = 0.2^\circ$ as shown in Fig.~\ref{fig:bias}.

\begin{figure}[tbp]
	\centering
	\vspace{0.5em}
	\flushleft
	\subfigure[Elevation estimation]{
		\begin{minipage}[b]{0.4\textwidth}
			\includegraphics[height=3.8in,width=5.2in]{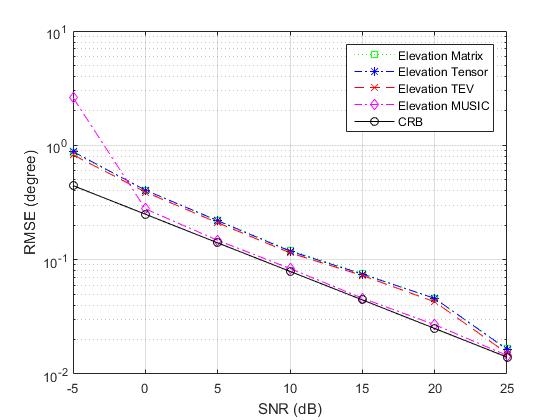}
			
		\end{minipage}
	} \\
	\subfigure[Azimuth estimation]{
		\begin{minipage}[b]{0.4\textwidth}
			\includegraphics[height=3.8in,width=5.2in]{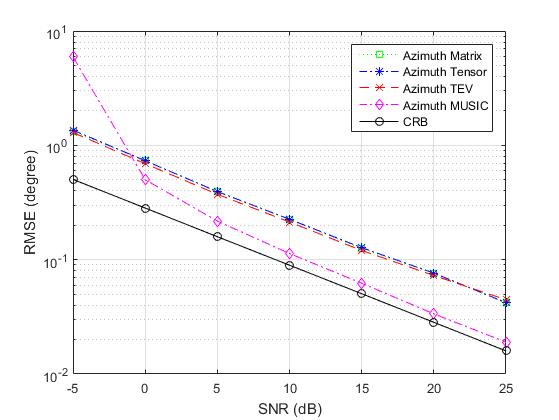}
			
		\end{minipage}
	}
	\caption{Elevation and azimuth estimation RMSE performance versus SNR of the matrix- and tensor-based ESPRIT methods, spectral MUSIC method, and the corresponding CRBs with $Q=6$, $K=1$, $L=32$ and $\Delta=0.1$. }
	\label{fig:s1q6}
\end{figure}

\begin{figure}[tbp]
	\centering
	\vspace{0.5em}
	\flushleft
	\subfigure[Elevation estimation]{
		\begin{minipage}[b]{0.4\textwidth}
			\includegraphics[height=3.8in,width=5.2in]{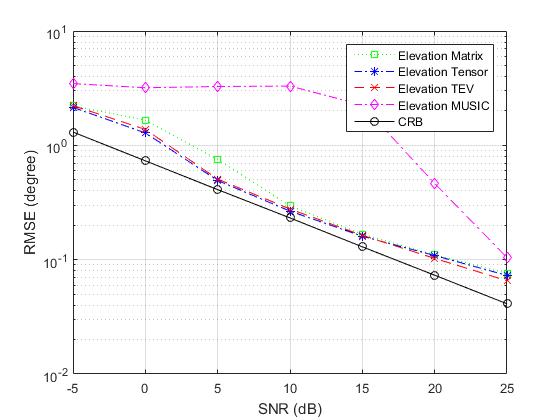}
			
		\end{minipage}
	} \\
	\subfigure[Azimuth estimation]{
		\begin{minipage}[b]{0.4\textwidth}
			\includegraphics[height=3.8in,width=5.2in]{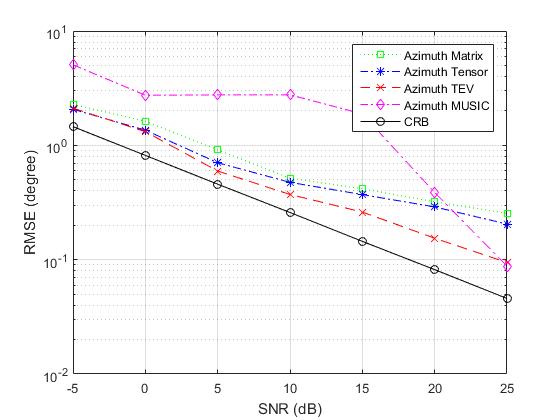}
			
		\end{minipage}
	}
	\caption{Elevation and azimuth estimation RMSE performance versus SNR of the matrix- and tensor-based ESPRIT methods, spectral MUSIC method and the corresponding CRBs with $Q=8$, $K=2$, $L=32$ and $\Delta=0.1$. }
	\label{fig:q8}
\end{figure}

By using the look-up table, the DOA estimation error can be reduced, especially in the high SNR region. Therefore, in our third example, we examine the elevation and azimuth RMSE performance for the proposed methods when the look-up table is also used for both single and multiple signal cases. The transmit array has 64 co-located antenna elements. The size of the virtual URA is $4\times 4$ with half wavelength displacement. The spatial sector of interest is set as $\Theta=[30^\circ, \, 40^\circ]$ and $\Phi=[65^\circ, \, 75^\circ]$. For the single target case, the DOA is set as $\phi = 66^\circ$ and $\theta = 34^\circ$, while for the multiple targets case, the DOAs are set as $\bm{\theta} = \{ 33^\circ, \, 39^\circ \}$, $\bm{\phi}= \{ 66^\circ, \, 71^\circ \}$.  The corresponding RMSEs are plotted versus SNR in Figs.~\ref{fig:s1q6} and \ref{fig:q8}, respectively, for the matrix- and tensor-based ESPRIT and 2D spectral MUSIC methods. Their corresponding CRBs are also plotted in the figures. It can be seen in Fig.~\ref{fig:s1q6} that the matrix- and tensor-based ESPRIT methods have almost the same elevation and azimuth RMSE performance. In comparison, the 2D spectral MUSIC method has a better elevation and azimuth RMSE performance and it achieves CRB because it is independent on the interpolation errors. However, the 2D spectral MUSIC requires an exhaustive 2D spectral search over the whole spatial region, which has a significantly higher computational complexity than the matrix- and tensor-based ESPRIT. In the multiple signal case, it can be observed in Fig.~\ref{fig:q8}(a) that the HOSVD-based ESPRIT and TEV-based method have slightly better elevation RMSE performance than that of the matrix-based ESPRIT method at SNR $<$10dB. Moreover, all three methods have almost the same RMSE performance above SNR $>$10dB. In addition, as can be seen in Fig.~\ref{fig:q8}(b), all three methods have almost the same azimuth RMSE performance at SNR $<$10dB, TEV-based method has a much better azimuth RMSE performance than the other two methods when SNR $>$10dB, because it is least influenced by the interpolation error as we discussed in Subsection~\ref{sec:bias} and shown in Fig.~\ref{fig:bias}. Moreover, the 2D spectral MUSIC method suffers from the small samples threshold effect. As a result, the 2D spectral MUSIC method shown poor elevation and azimuth RMSE performance, which gets closer to the one by the ESPRIT methods only for SNR$>$25dB and SNR$>$20dB for elevation and azimuth, respectively.

\begin{figure}[tbp]
	\centering
	\vspace{0.5em}
	\flushleft
	\subfigure[Elevation resolution ability]{
		\begin{minipage}[b]{0.4\textwidth}
			\includegraphics[height=3.8in,width=5.2in]{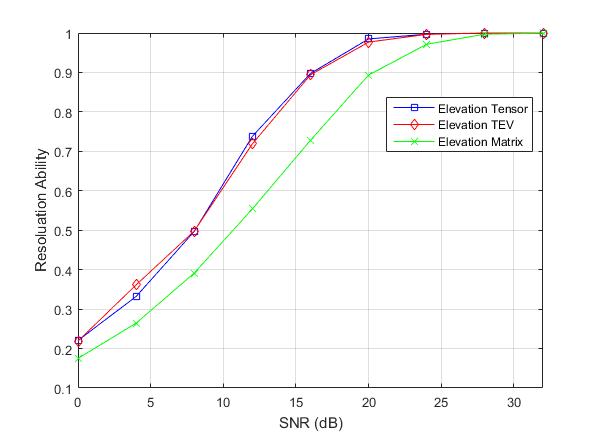}
			
		\end{minipage}
	} \\
	\subfigure[Azimuth resolution ability]{
		\begin{minipage}[b]{0.4\textwidth}
			\includegraphics[height=3.8in,width=5.2in]{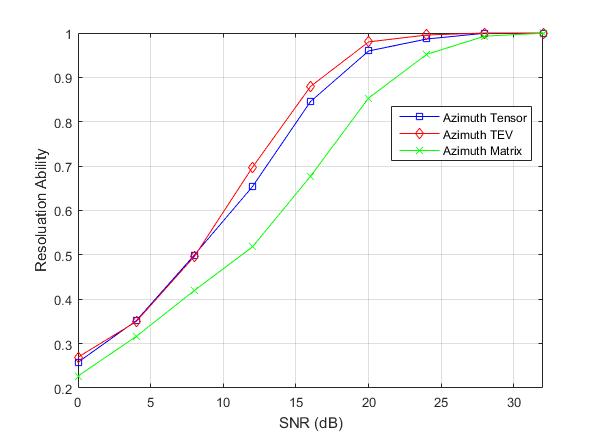}
			
		\end{minipage}
	}\caption{Resolution ability of two closely located targets with virtual transmit URA.}
	\label{fig:resolution}
\end{figure}

In the last experiment, we investigate the resolution ability of the 2D TB and interpolation MIMO radar with matrix- and tensor-based ESPRIT methods. The elevation and azimuth of two targets are set as $\bm{\theta} = \{ 36^\circ, 39^\circ \}$ and $\bm{\phi } = \{ 66^\circ, 69^\circ \}$, respectively.  The transmit array has 64 co-located antenna elements. The size of the virtual URA is $4\times 4$.  The spatial sector of interest is set as $\Theta=[30^\circ,40^\circ]$, $\Phi=[65^\circ,75^\circ]$.  The targets are considered as resolvable if the following condition is satisfied for the elevation \cite{Trees02}
\begin{equation}
	| \hat{\theta}_p - \theta_p | < \frac{ |\theta_1 - \theta_2| }{2} \label{eq:resoltion}
\end{equation}
where $\hat{\theta}_p, p=1,2$ represent the estimated elevation parameters of two targets. This criterion is also suitable for testing the azimuth resolution ability. Fig. \ref{fig:resolution} shows that the resolution probability vesus SNR for matrix- and tensor-based ESPRIT methods. It can be seen that both methods achieve 100\% correct resolution at the SNR of 25dB for elevation and at the SNR of 27dB for azimuth. Also, the tensor-based ESPRIT method has a higher resolution ability than that of the matrix-based method when SNR $<$25dB.

\section{Conclusion}
The problem of the 2D transmit array interpolation and beamspace design for mono-static MIMO radar with application to elevation and azimuth estimation has been addressed in this paper.
The 2D transmit array interpolation and beamspace design problem has been formulated, for example, as the minimax convex optimization problem with constraints on array interpolation errors within a spatial sector of interest while minimizing the transmit power outside the sector. The desired structure of the virtual transmit array is also enforced. The virtual transmit array can be of any desirable shape, but practically most appealing shapes are the L-shaped array and URA because they allow to benefit from translational invariance property when estimating elevation and azimuth parameter at the receiver. By taking advantage of the high-dimensional structure inherent in the received data, the proposed MIMO radar model has been folded into a higher-order tensor. Then tensor-based ESPRIT methods with direct data approach have been devised. Furthermore, we have analysed the DOA estimation bias caused by transmit array interpolation errors and have built an offline look-up table aiming to decrease the DOA estimation bias. The corresponding deterministic CRB on DOA estimation in the proposed 2D TB and interpolation MIMO radar has been also provided. Simulation results have verified the effectiveness of the proposed methods.

\appendix[]
\subsection{Interpolation Errors Analysis}
We derive the MSE of $\mu_k$ as an example. In the absence of noise, the solution of \eqref{eq:hosvdele} can be expressed with the matrix-based signal subspace as
\begin{equation}
	\bm{\Theta}= \left( \bm{\Omega}_{11}\bm{U}_s \right)^\dagger \bm{\Omega}_{21} \bm{U}_s.
\end{equation}
If there exists interpolation errors, we have
\begin{equation}
	(\bm{\Omega}_{11}\bm{U}_s+\Delta\bm{U}_{s1})(\bm{\Theta}+\Delta\bm{\Theta})=\bm{\Omega}_{21}\bm{U}_s+\Delta\bm{U}_{s2}.
\end{equation}
Neglecting the cross item, $\Delta\bm{\Theta}$ can be calculated as
\begin{equation}
	\Delta\bm{\Theta}=\bm{U}_s^\dagger(\Delta\bm{U}_{s2}-\Delta\bm{U}_{s1}\bm{\Theta}) \label{eq:subper}
\end{equation}
where
\begin{align}
\Delta\bm{U}_{s1}=(\bm{\Omega}_{11}\bm{U}_s)^\dagger(\bm{\Omega}_{21}\bm{U}_n)\bm{U}_n^H\Delta\bm{A}(\theta,\phi)\bm{A}^{\ddagger} (\theta,\phi) \bm{U}_s \\
\Delta\bm{U}_{s2}=(\bm{\Omega}_{11}\bm{U}_s)^\dagger(\bm{\Omega}_{11}\bm{U}_n)\bm{U}_n^H\Delta\bm{A}(\theta,\phi)\bm{A}^{\ddagger} (\theta,\phi) \bm{U}_s.
\end{align}
The first-order perturbation of $k$th eigenvector of $\bm{\Theta}$ is therefore given as
\begin{equation}
	\Delta\lambda_k=\bm{p}_k^T\Delta\bm{\Theta}\bm{q}_k. \label{eq:eigper}
\end{equation}
Moreover, the relationship between $\Delta\mu_k$ and $\Delta\lambda_k$ can be obtained as
\begin{equation}
	\Delta\mu_k=c_k\text{Im}\left\{ \frac{\Delta\lambda_k}{\lambda_k} \right\} \label{eq:deltamu}
\end{equation}
where $c_k$ depends on the array structure. Based on the angle-root relationship \cite{Tufts89}, we have $c_k=1/\pi$ for URA.    

Substituting \eqref{eq:eigper} into \eqref{eq:deltamu}, we finally obtain
\begin{equation}
\Delta\mu_k=\frac{1}{\pi}\text{Im} \left\{ \bm{\alpha}_k^H \Delta\bm{A} (\theta,\phi) \bm{\beta}_k \right\}.
\end{equation}

\subsection{Conversion of Direction Cosine Functions to DOA Estimation Biases}

The first-order partial derivatives of $\theta$ and $\phi$ with respect to $\mu$ and $\nu$ for URA are 
\begin{align}
	&\frac{\pa\theta}{\pa\mu} = \frac{\mu}{\sqrt{\mu^2+\nu^2}},  \quad \frac{\pa\theta}{\pa\nu} = \frac{\nu}{\sqrt{\mu^2+\nu^2}} \label{eq:first order1} \\
	&\frac{\pa\phi}{\pa\mu} = - \frac{\nu}{\mu^2 + \nu^2}, \quad \frac{\pa\phi}{\pa\nu} = \frac{\mu}{\mu^2+\nu^2}. \label{eq:first order2}
\end{align}

The multi-parameters Taylor series expansion formula is given as
\begin{equation}
	f(x_0+h,y_0+k)\approx f(x_0+y_0)+h\frac{\pa f(x_0,y_0)}{\pa x}+k\frac{\pa f(x_0,y_0)}{\pa y} \label{eq:taylor}
\end{equation}

Based on \eqref{eq:taylor}, we can easily obtain the approximation of $\hat{\theta}$ and $\hat{\phi}$ around, respectively, $\theta$ and $\phi$ as
\begin{align}
	&\hat{\theta} \approx \theta + \frac{\pa\theta}{\pa\mu} \Delta \mu + \frac{\pa\theta}{\pa\nu} \Delta\nu \label{eq:taylor expansion1} \\
	&\hat{\phi} \approx \phi + \frac{\pa\phi}{\pa\mu} \Delta \mu + \frac{\pa\phi}{\pa\nu} \Delta \nu	\label{eq:taylor expansion2}
\end{align}

Substituting \eqref{eq:first order1} and \eqref{eq:first order2} into \eqref{eq:taylor expansion1} and \eqref{eq:taylor expansion2}, respectively, and neglecting the second-order and cross terms, we finally find the relationships between the perturbation of direction cosine functions and DOA estimation biases.

\end{document}